\documentclass{JINST}
\usepackage{amsmath}
\usepackage{txfonts}
\usepackage{latexsym}
\usepackage[numbers]{natbib}

\newcommand{\Planck}{\textsc{Planck}}

\title{Off-line radiometric analysis of \Planck/LFI data} 

\author{M.~Tomasi, A.~Mennella, S.~Galeotta, S.R.~Lowe, L.~Mendes, R.~Leonardi, F.~Villa, B.~Cappellini, A.~Gregorio, P.~Meinhold, M.~Sandri, F.~Cuttaia, L.~Terenzi, M.~Maris, L.~Valenziano, M.~J.~Salmon, M.~Bersanelli, P.~Binko, R.~C.~Butler, O.~D'Arcangelo, S.~Fogliani, M.~Frailis, E.~Franceschi, F.~Gasparo, G.~Maggio, D.~Maino, M.~Malaspina, N.~Mandolesi, P.~Manzato, M.~Meharga, G.~Morgante, N.~Morisset, F.~Pasian, F.~Perrotta, R.~Rohlfs, M.~T\"urler, A.~Zacchei, A.~Zonca.}

\abstract{The \Planck{} Low Frequency Instrument (LFI) is an array of 22 pseudo-correlation radiometers on-board the \Planck{} satellite to measure temperature and polarization anisotropies in the Cosmic Microwave Background (CMB) in three frequency bands (30, 44 and 70\,GHz).\\ To calibrate and verify the performances of the LFI, a software suite named LIFE has been developed. Its aims are to provide a common platform to use for analyzing the results of the tests performed on the single components of the instrument (RCAs, Radiometric Chain Assemblies) and on the integrated Radiometric Array Assembly (RAA). Moreover, its analysis tools are designed to be used during the flight as well to produce periodic reports on the status of the instrument.\\ The LIFE suite has been developed using a multi-layered, cross-platform approach. It implements a number of analysis modules written in RSI IDL, each accessing the data through a portable and heavily optimized library of functions written in C and C++. One of the most important features of LIFE is its ability to run the same data analysis codes both using ground test data and real flight data as input.\\ The LIFE software suite has been successfully used during the RCA/RAA tests and the \Planck{} Integrated System Tests. Moreover, the software has also passed the verification for its in-flight use during the System Operations Verification Tests, held in October 2008.}

\keywords{Cosmic microwave background - Methods: data analysis - Methods: numerical}

\begin{document}

\section{Introduction}

\begin{figure*}
  \begin{center}
    \begin{tabular}{cc}
      \includegraphics[height=5.5cm]{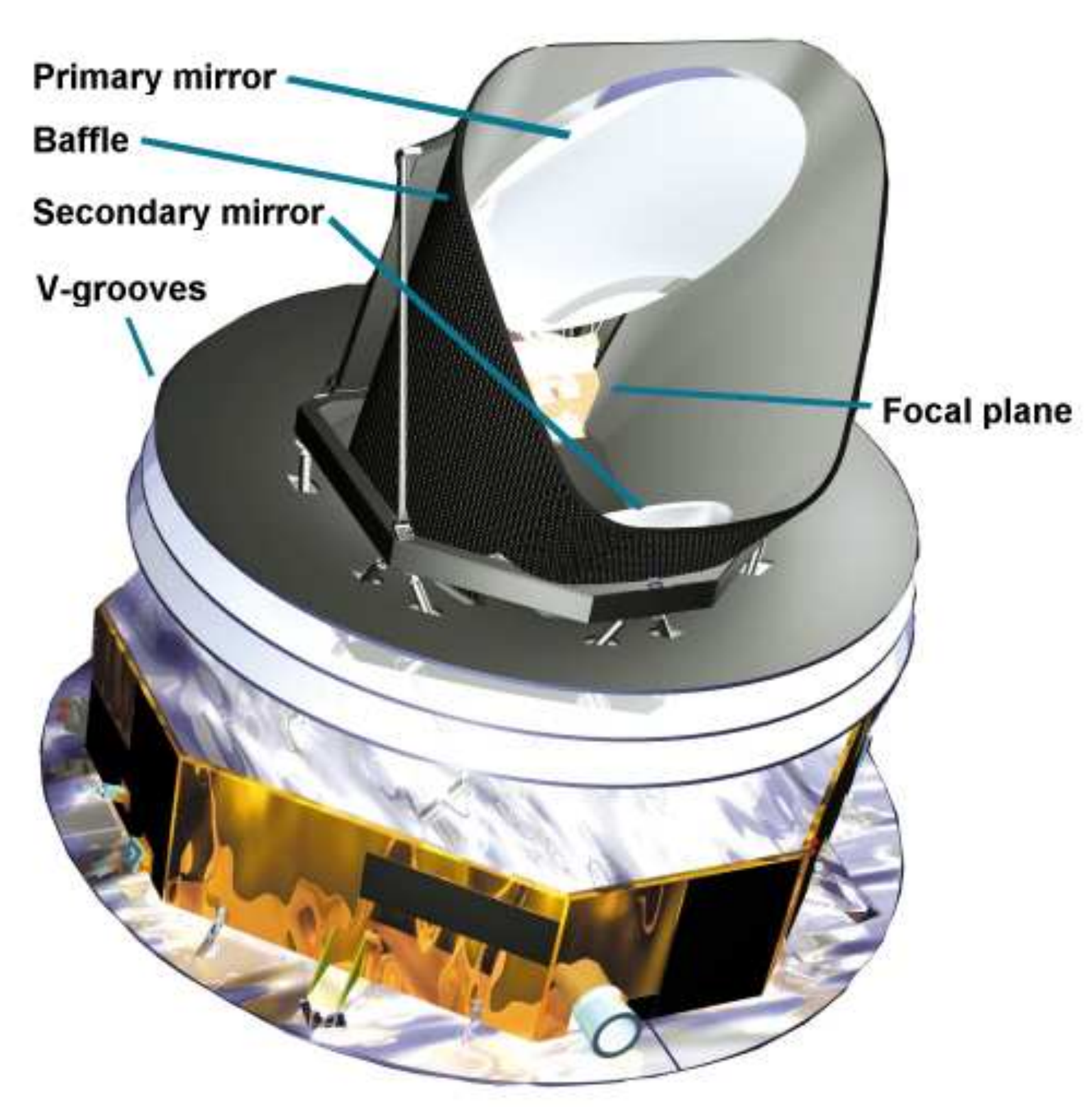} &
      \includegraphics[height=5.5cm]{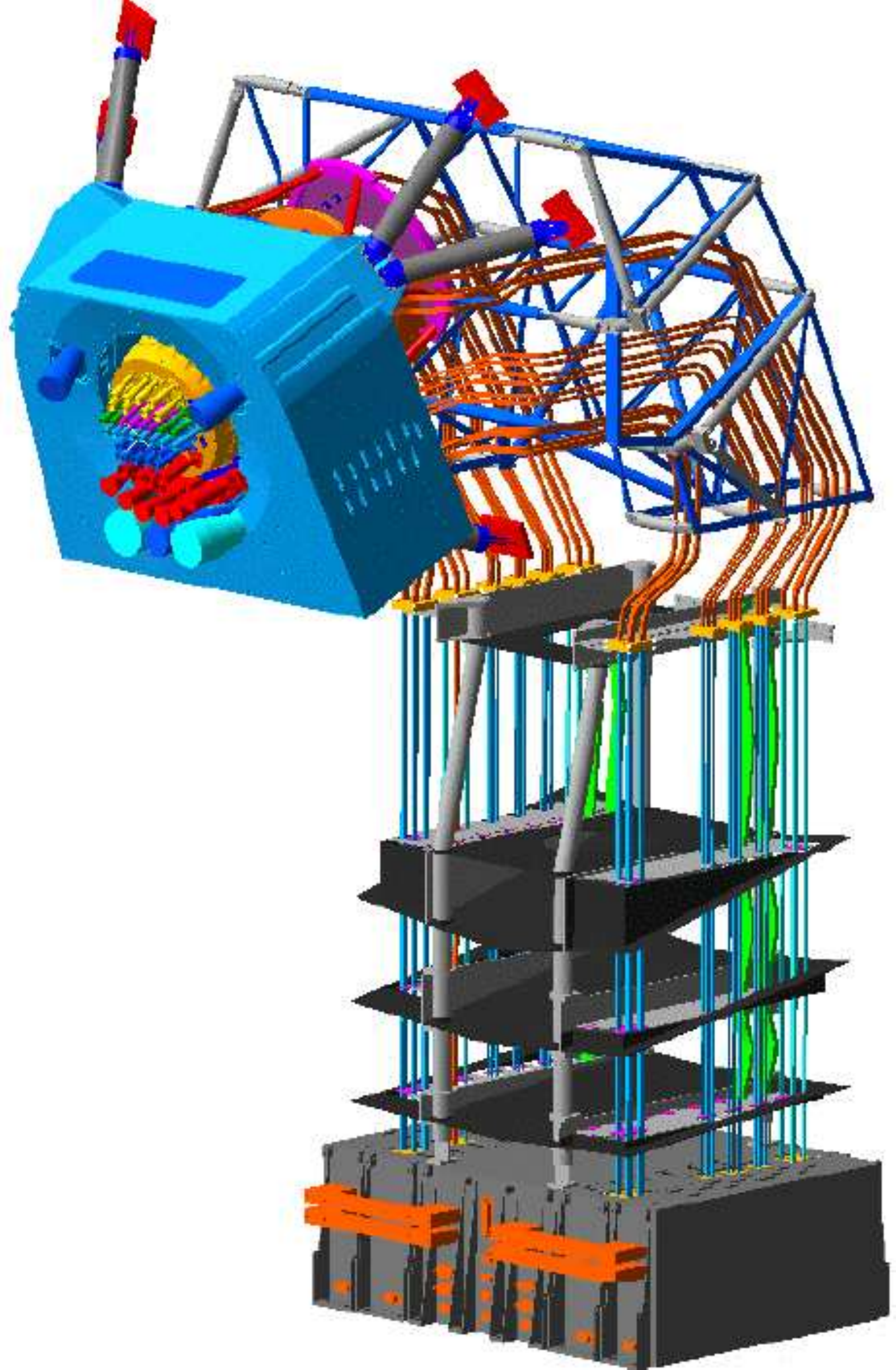} \\
    \end{tabular}
  \end{center}

  \caption{\label{fig:planck} \textbf{Left}: schematics of the \Planck{} satellite. The warm service module ($\sim$ 300\,K) at the bottom of the satellite is decoupled from the focal plane and the telescope by means of three thermal radiators called ``V-grooves''. The LFI focal plane is cooled to 20\,K by a hydrogen Sorption Cooler (SC) which acts also as a pre-stage for the HFI 4\,K cooler. \textbf{Right}: detailed view of the LFI structure (the so-called RAA, Radiometric Array Assembly). On top, the cold Focal Plane Unit (FPU) with the LFI and HFI feed horns is shown. A set of waveguides connects the FPU with the warm (300\,K) Back End Unit (BEU), shown at the bottom.}
\end{figure*}

The Low Frequency Instrument (LFI) on-board the ESA \Planck{} space mission (figure \ref{fig:planck}) is an array of 22 pseudo-correlation differential radiometers cryogenically cooled at 20\,K \cite{2009_LFI_cal_M1,2009_LFI_cal_M2}. It will measure the temperature and polarization anisotropies of the Cosmic Microwave Background (CMB) in the 30--70\,GHz range with an angular resolution of 14'--33' and a sensitivity of a few $\mu$K per pixel in the final maps.

The LFI has undergone a comprehensive test campaign, where the instrument has been verified and calibrated at different integration stages \cite{2009_LFI_cal_M3,2009_LFI_cal_M4}. To analyze the data collected during the tests, part of the \Planck{}/LFI instrument team developed the LIFE software suite, which is composed of the three following modules:

\begin{enumerate}
\item RaNA (Radiometric aNAlyser), used to test each Radiometer Chain Assembly (RCA) before the integration in the LFI.

\item LAMA (LFI Array Measurement Analyser), used to test the integrated instrument (RAA -- Radiometer Array Assembly).

\item Pegaso, a tool that will be used for the verification and calibration of LFI during flight. Pegaso has already been used successfully during the Planck System Operations Verification Tests (SOVT) in October, 2008.
\end{enumerate}

This paper focuses primarly on RaNA and LAMA, discussing their development, implementation and use during the LFI RCA/RAA tests. A brief description of Pegaso is provided in section \ref{sec:Pegaso}.

The outline of this article is the following. In sect.~\ref{sec:PlanckLFI} we provide an overview of LFI and the RCA/RAA
test campaign. In sect.~\ref{sec:LIFE} we present the LIFE analysis tool: sect.~\ref{sec:LIFEOverview} provides a high-level overview, while details about RaNA and LAMA are discussed in sect.~\ref{sec:RaNA} and sect.~\ref{sec:LAMA} respectively. In sect.~\ref{sec:Pegaso} we explain how LIFE is going to be used during flight operations. Finally, sect.~\ref{sec:Conclusions} reports our conclusions about this work.


\section{The \Planck{}/LFI Instrument}
\label{sec:PlanckLFI}

\subsection{Overview of the Instrument}

\begin{figure*}[tf]
  \centering
  \includegraphics[width=0.8\textwidth]{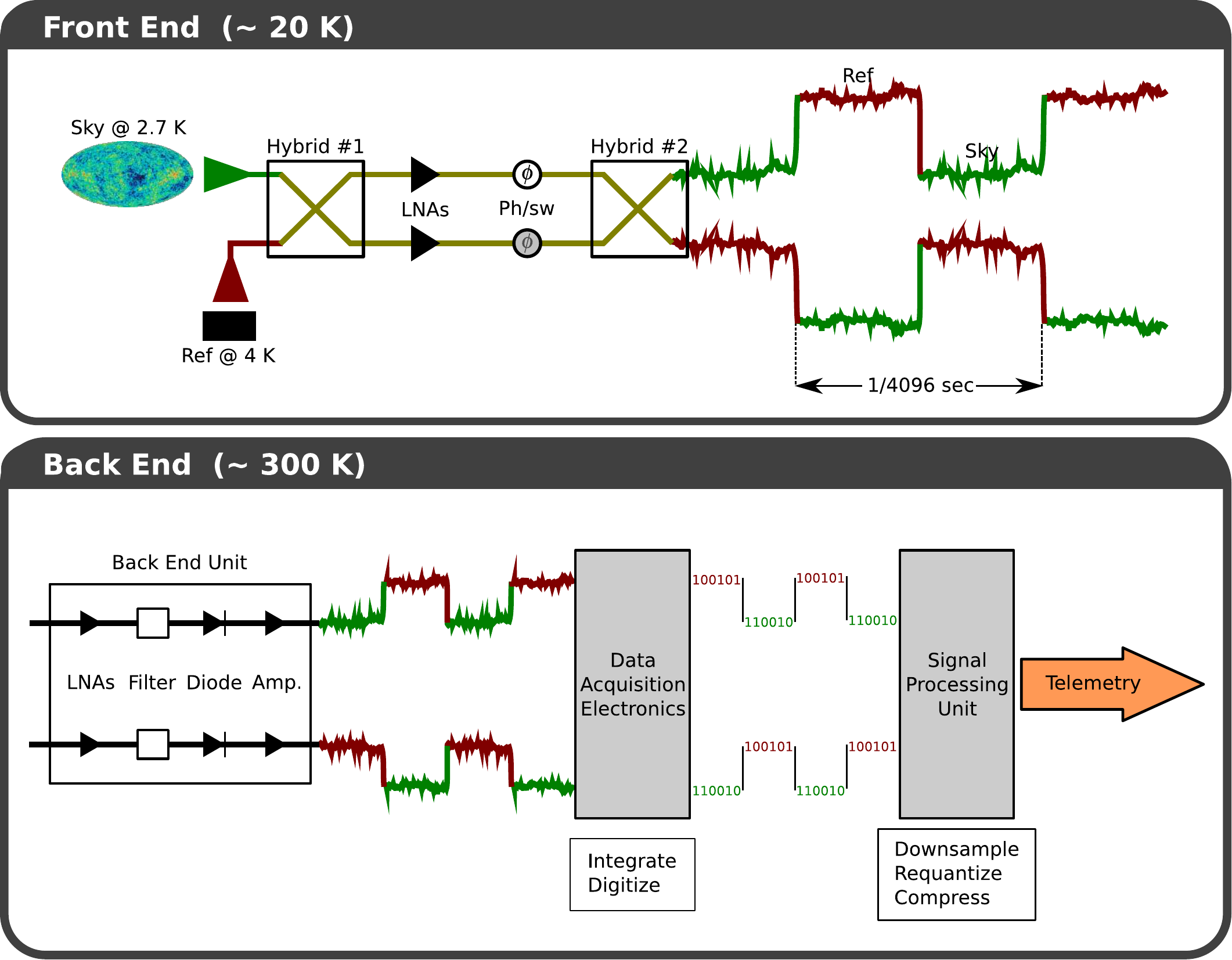}
  \caption{Schematics of an LFI radiometer. \textbf{Top}: in the
    front-end of each radiometer (cooled at $\sim$20\,K) the polarized
    signal of the sky coming from an orthomode transducer (not shown)
    is combined with the stable signal of a reference load at $\sim$4\,K.
    The two signals are mixed together, amplified and switched
    with a frequency of 4096\,Hz by a phase switch. \textbf{Bottom}:
    through a set of waveguides the signal enters the warm back end
    ($\sim$300\,K), where the signal is detected by a diode and
    further amplified. Two modules (the DAE, Data Acquisition
    Electronics, and the SPU, Signal Processing Unit) integrate,
    digitize, mix and compress the signal, which is then transmitted to
    Earth.}
  \label{fig:RCASchema}
\end{figure*}

Each LFI receiver performs a differential measurement of the sky signal ($\sim$ 2.7\,K) by comparing it with the signal of a stable reference load made from Eccosorb\footnote{\href{http://www.emersoncuming.com}{http://www.emersoncuming.com}.} and kept at a temperature of $\sim$ 4.5\,K \cite{2009_LFI_cal_M2,mennella1f}. Due to the differential nature of each receiver, the output of each one is detected through two distinct channels. The output of LFI is therefore represented by the data produced by 44 channels, each alternatively detecting both the sky and the reference signals. Refer to fig.~\ref{fig:RCASchema} for further details.

\subsection{The LFI Test Campaign}

Before integration into the satellite, LFI has been tested through a number of phases, each of them working at a different integration level. The LFI test sequence has been designed with the following objectives \cite{LFICalibrationPlan}:
\begin{enumerate}
\item to check the correct operation of every critical part of the radiometers;

\item to measure those quantities whose knowledge is needed for data analysis (e.g. white noise level, $1/f$ knee frequency);

\item to calibrate the instrument and to perform susceptibility tests (e.g. impact of temperature fluctuations on the output signal).
\end{enumerate}

In this article we concentrate on the so-called RCA and RAA tests of LFI. The RCA is a system composed of a single feed horn and two equal radiometers measuring the polarized components of the signal coming from the Planck telescope. In the RAA (Radiometer Array Assembly) tests the whole LFI -- with all the RCAs integrated -- has been tested. The Alcatel/Alenia Space laboratories in Vimodrone, Milan (Italy) have hosted the RCA tests for the 30 and 44\,GHz chains and the tests on the RAA for both qualification (QM) and flight (FM) models. The 70\,GHz chains have been tested in the Millilab laboratories (Finland).

The LIFE data analysis software has been developed in parallel with the test campaign and has been used throughout the test phases up to the satellite-level tests.


\section{Calibration and verification of LFI using LIFE}
\label{sec:LIFE}

\subsection{Motivation and requirements of a dedicated analysis tool}
\label{sec:LIFEMotivations}

The calibration campaign of the LFI has been a challenging and complex task, involving several tests which have often been custom designed for each specific purpose. It was therefore recognized since the beginning that a dedicated data analysis tool was necessary, with the following requirements:
\begin{enumerate}
\item The tool must allow the analysis of the data produced during the overall flight RCA and RAA test campaigns.

\item It should support the analysis of all the RCA/RAA tests foreseen for the \Planck{}/LFI.

\item It must be fast enough to perform near real-time analysis when needed (\emph{true} real-time analysis is not needed).

\item It should be based mostly on open-source tools, using proprietary software only when really needed. (This allows full access to the source code whenever some problem with the software arises.)

\item The analysis tools must be portable among all the most important platforms (Windows, UNIX, Mac OS X) used by the \Planck{}/LFI scientific team.

\item Because of its widespread usage in the \Planck{}/LFI community, it was required to allow the development of data analysis code in ITT IDL\footnote{\href{http://www.ittvis.com/ProductServices/IDL.aspx}{http://www.ittvis.com/ProductServices/IDL.aspx}. An alternative would have been Python and one of its numerous numerical libraries (e.g.\ NumPy), which have the advantage of being open source products. However, very few people in our collaboration knew how to program in Python when we begun developing LIFE in 2004, and open source libraries available were not as numerous as they are today (e.g.\ PyQt for Windows GPL would have been released only one year later).}.
\end{enumerate}

Because all these requirements could not be fulfilled by market-available software, a dedicated suite was developed: LIFE (Lfi Integrated perFormance Evaluator).

\subsection{Structure of LIFE}
\label{sec:LIFEOverview}

The purpose of LIFE is to provide a common tool to perform all the main analysis tasks during the RCA/RAA campaign, and to allow the usage of the same analysis tools during flight operations as well.

LIFE has been designed mainly to work off-line. In each RCA/RAA test, the output and status of the component under study was recorded by the acquisition system \cite{2004ZaccheiGroundTestData,2009_LFI_cal_D1}. After the end of the test, the acquisition was stopped and the collected data made available to the LFI scientific team members, who would then analyse them using LIFE on their own computers. Dedicated computers with LIFE installed were also available in the laboratory --- this was particularly useful for those tasks that needed a quasi real-time analysis, like some of the tuning tests at the beginning of the calibration campaign \cite{2009_LFI_cal_R7}.

\begin{figure}[tbf]
  \centering
  \includegraphics[width=7cm]{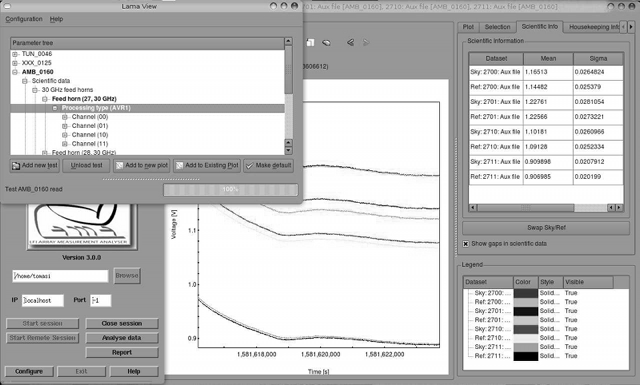}
  \caption{Screenshot of LAMA. On the top left, the Lama View window
    shows the list of tests and parameters that can be plotted or
    analysed. On the bottom left there is the Lama main window. In the
    background, a Plot window is opened, showing the output of the
    four channels of a radiometer.}
  \label{fig:LAMA}
\end{figure}

We implemented LIFE according to a modular structure, as a suite of three independent analysis tools able to run under different environments. Each environment acts like a data provider for the analysis modules and at the same time provides the user with a consistent graphical user interface to navigate and select the data to be used for the analysis (see figure \ref{fig:LAMA}). The three environments in the suite are:
\begin{description}
\item[RaNA] (Radiometer aNAlyser) has been used during the RCA tests (both the tests done on the 30-44 GHz chains and on the 70 GHz ones).
\item[LAMA] (LFI Array Measurements Analyser) has been used during the RAA tests.
\item[Pegaso] is the environment to be used during flight. Refer to sect.~\ref{sec:Pegaso}.
\end{description}
All the analysis codes (with few exceptions) can be run within all the environments.

Due to the large number of members of the Planck team developing analysis modules, LIFE was designed to be easily expandable; furthermore, IDL was chosen as the main language to implement the data analysis modules. Adopting IDL (together with a centralized CVS system to keep track of any code change) has allowed potentially any member of the LFI scientific team to examine and improve the LIFE analysis code.

The I/O part of the code has been developed using C and C++ (mainly via the CFITSIO library\footnote{\href{http://heasarc.gsfc.nasa.gov/docs/software/fitsio/}{http://heasarc.gsfc.nasa.gov/docs/software/fitsio/}}) in order to optimize data reading and writing speed, considering the large volume of data produced during the RCA and RAA tests \cite{2009_LFI_cal_D3}:
\begin{enumerate}
\item During the RCA test campaign, the radiometric output was sampled at the full frequency of the LFI electronics (8192\,Hz for each of the four RCA legs), leading to a data rate of $\sim 200$\,MB/hour.

\item During the RAA test campaign, the output was sampled at a sampling rate ranging from $\sim 32$\,Hz at 30\,GHz to $\sim 77$\,Hz at 70\,GHz. Considering also several hundreds of housekeeping parameters sampled at 1\,Hz, the overall data rate was similar to that of a single RCA at full sampling rate.
\end{enumerate}

In order to improve their responsiveness, both RaNA and LAMA employed subsampling techniques to reduce the amount of radiometric data to process. In fact, the analysis for a number of RCA/RAA tests only requires to know a few statistics of the radiometric output\footnote{E.g.\ tuning the LNAs only requires the knowledge of how the average voltage output of the radiometer varies while changing the LNA biases \cite{2009_LFI_cal_R7}. Also, downsampled data has been systematically used in the analysis of very long RCA/RAA tests (several hours) in order to determine the time when some particular event happened. Once a temporal window for that event has been established, LIFE was used to retrieve the full radiometric data within that window only.}. Therefore, LIFE provides the ability to use a set of auxiliary FITS files
(conventionally called AUX files), each of them containing a $4\times N$ matrix where each row is of the form $(t, \bar V, \sigma, N)$: $t$ is the time in seconds and $\bar V$, $\sigma$, $N$ are the average radiometric output, standard deviation and number of raw\footnote{Ideally this number should be constant (e.g.\ for RCA tests $N = 4096$ for both sky and reference data streams) for each AUX sample, but data losses in the connection between the instrument and the data acquisition board can lead to a reduction.} samples within the time interval $[t, t + 1\,\mathrm{s}]$. These four parameters allow LIFE to straightforwardly reconstruct the correct average and standard deviation of the radiometric signal over an arbitrary time range $[t_0, t_1]$, where both $t_0$ and $t_1$ have one-second resolution (see app.~\ref{sec:downsampledMath}). On the other side, LIFE is able to retrieve the full radiometric data when needed.

The advantage of AUX files lies in their compactness and manageability, their size being a few orders of magnitude smaller than the FITS files containing the whole radiometric output. This allows a user to open and navigate in a test spanning several hours in a few seconds.

\subsection{RCA Data Analysis using RaNA}
\label{sec:RaNA}

\begin{figure}[tbf]
  \centering
  \includegraphics[width=\columnwidth]{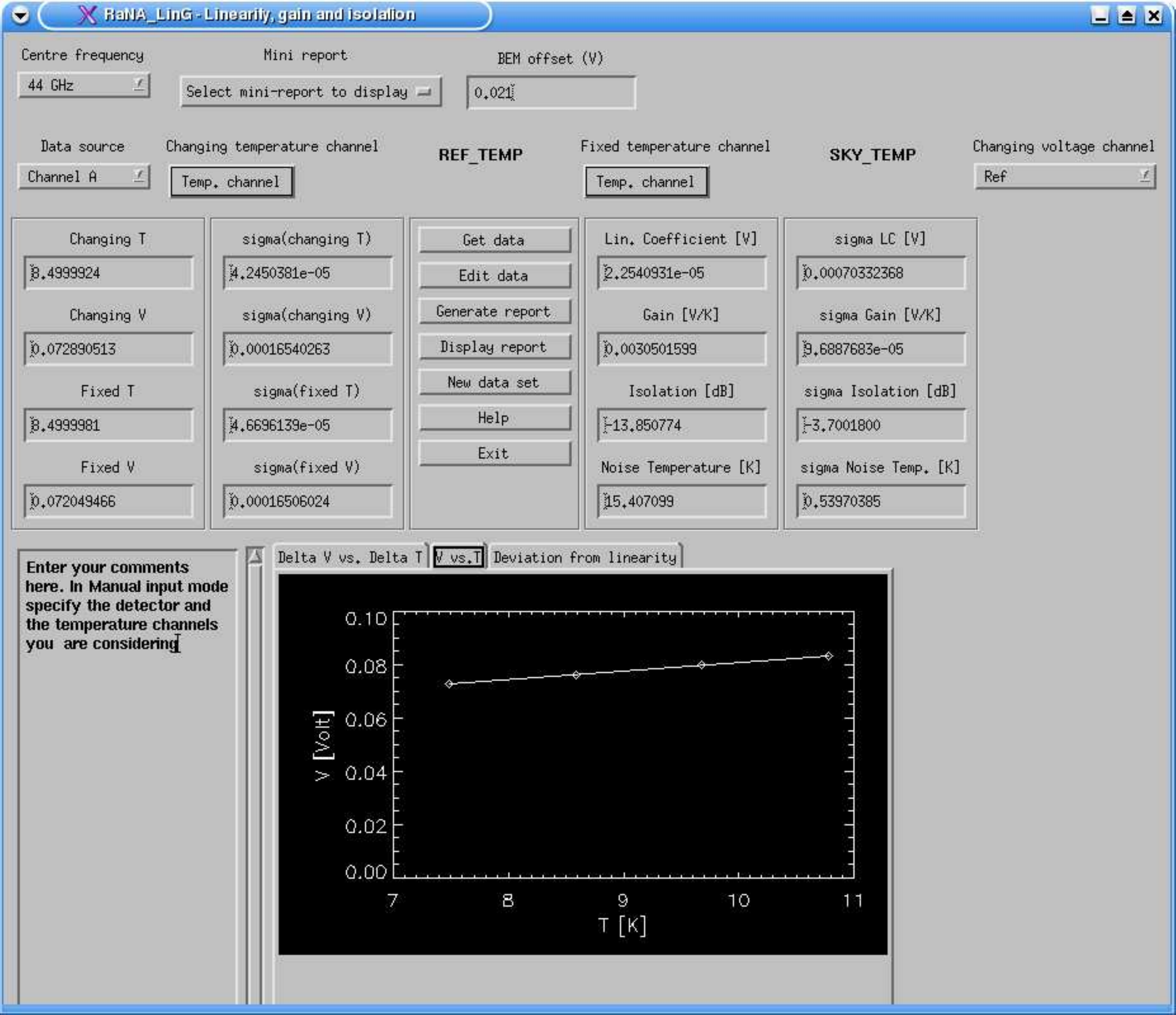}
  \caption{Screenshot of one of the RaNA analysis modules, LinG. The module is used to determine the photometric calibration constant for the LFI radiometers, as well as their noise temperature. Input data (such as average input levels) can be either entered manually or automatically retrieved from test data. As for every LIFE analysis module, the results of the calculation and the plots can be saved into a \LaTeX{} report by pressing the ``Generate report'' button.}
  \label{fig:LinG}
\end{figure}

RaNA was the first LIFE environment to be developed. It is a tool to calibrate and validate the LFI QM/FM RCAs \cite{2009_LFI_cal_M4} and implements a number of analysis modules. In the following paragraphs we are going to illustrate its fundamentals.

\subsubsection{The Data Access Library}

RaNA provides access to the test data through a set of data access functions written in IDL. These functions are available from the IDL command line as well. This allows the user to use the power of IDL to perform some quick calculations and draw plots interactively. For instance, the following IDL command will interpolate the output of the first channel (index 0) of the RCA under study with a straight line:

\begin{verbatim}
print, poly_fit (rana_get_sky_x (0), 
                 rana_get_sky_y (0), 1)
\end{verbatim}
The purpose of \texttt{rana\_get\_sky\_x~(0)} and \texttt{rana\_get\_sky\_y~(0)} is to retrieve the sky stream from the
radiometric output of channel 0 (out of the 4 channels in each RCA) as X (time) and Y (voltage) coordinates respectively. Similar functions exist for accessing the reference load stream (e.g.\ \texttt{rana\_get\_ref\_x}) and the scientific selection (e.g.\ \texttt{rana\_selection\_get\_sky\_x}).

\subsubsection{The Analysis Modules}

RaNA implements a number of analysis modules: a few examples of such modules are the ones used for the tuning of the front-end phase switches and amplifiers, and a module that produces spectrograms from time-ordered data series. Each RaNA analysis module implements a command-line interface and can be used within IDL scripts. Complex analysis modules can be therefore built using simpler modules.

A number of analysis modules provide a Graphical User Interface (GUI) to enter the data needed for the calculation (either manually or by automatically retrieving them from the test data) and to show the output of the module. Currently, the modules providing a full GUI are the following:
\begin{description}
\item[LinG:] estimation of the linearity and gain of the four RCA channels. The module also has the ability to estimate the noise temperature.

\item[FFT:] estimation of the signal noise properties (e.g.\ white noise level, $1/f$ knee frequency).

\item[Susc:] analysis of the susceptibility towards systematic effects (such as temperature fluctuations).

\item[SPR:] determination of the radiometric spectral response (bandpasses).
\end{description}

RaNA also implements a general-purpose module, RaNA View, for quick plotting of both radiometric and housekeeping data.

\subsection{RAA Data Analysis using LAMA}
\label{sec:LAMA}

\begin{figure}[tbf]
  \centering
  \includegraphics[width=9cm]{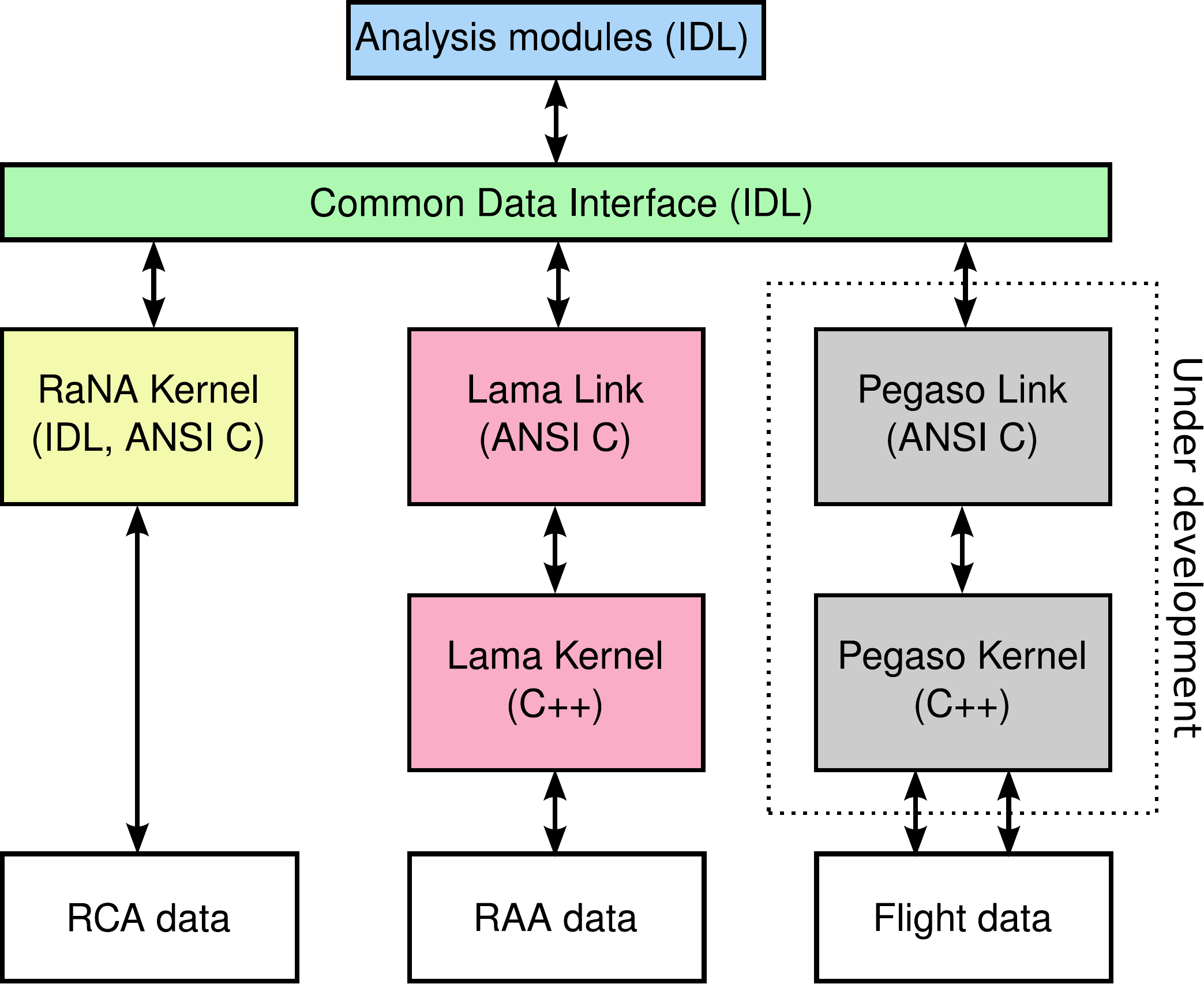}
  \caption{LIFE modules access data through a layered Application Program Interface (API). Each module uses a Common Data Interface to retrieve radiometric and housekeeping data. The context determines if either RCA or RAA data is needed, and thus if the data must be provided either by RaNA or LAMA. Internally LAMA uses sockets to exchange data between the LAMA kernel and the Common Data Interface. This is the same baseline being used to develop Pegaso. The two arrows connecting Pegaso with the flight data represent its ability to access both plain FITS files and the \Planck{} ground database (see sect.~\protect\ref{sec:Pegaso}).}
  \label{fig:dataAccess}
\end{figure}

The purpose of LAMA is to provide the scientific team with a tool to calibrate and validate the RAA QM/FM instruments \cite{2009_LFI_cal_M3}. Even if its purpose is similar to those of RaNA, there are some significant differences:

\begin{enumerate}
\item The data acquisition pipeline no longer uses the RCA acquisition system as described by \cite{2004ZaccheiGroundTestData}, but is the same one to be used during flight \cite{2003ZaccheiTelemetry}, which is considerably different.

\item The number of data channels to be handled by LAMA is larger than in the RCA case. This applies to both the number of radiometric channels and the number of housekeeping parameters (several hundred).
\end{enumerate}

Therefore, we had to develop a system that allows the analysis modules to access the data in a way that is independent of the data format. This system is called the Common Data Interface (CDI).

\subsubsection{The Common Data Interface}

LAMA implements a Common Data Interface (CDI) that allows LIFE analysis modules to access RAA data in the same way as they access RCA data under RaNA. Therefore, no radiometric analysis module that was already available within RaNA needed to be re-implemented to work with LAMA, despite the differences in the format and layout of the RCA/RAA data. The LIFE CDI provides a set of functions that accept as the first argument a string identifying the environment that shall provide the information, either \texttt{'rana'} or \texttt{'lama'}. Depending on the value of the string, the appropriate RaNA/LAMA function will be called.

It is possible to run the same analysis module under RaNA and LAMA at the same time, as every module is fully encapsulated within its environment (i.e.\ no global data structures are used). This is very useful e.g.\ to compare the results of an RCA test and an RAA test.

\subsubsection{Implementation of Speed-Critical Tasks}

The most speed-critical parts of LAMA are written in C++. These include the I/O routines and the Lama View module (the counterpart of RaNA View, see figure \ref{fig:LAMA}). This is because accessing RAA data I/O presents more caveats than RCA data:
\begin{enumerate}
\item The acquisition pipeline saves the output of each LFI detector and housekeeping parameter into separate files, for a total of $\sim 10^3$ datastreams;

\item In order to minimise chances of data loss, each parameter has its datastream split in chunks of one hour each and saved into separate files. Together with point 1, this means that for a typical test lasting a few hours, roughly $10^3$ -- $10^4$ files are produced. Prior to the true loading of the scientific and housekeeping data, LAMA must read and interpret the header of each FITS file in order to determine its contents.

\item Unlike the RCA data acquisition pipeline, the RAA pipeline does not create downsampled AUX files. Therefore, LAMA must downsample scientific data ``on the fly''.
\end{enumerate}
An additional requirement on the speed and memory footprint of LAMA came from the decision of allowing the program to load, plot and analyze multiple tests at the same time. This of course pushed the need for code optimization.

We therefore developed \emph{Lama View}, a stand-alone C++ application based on the Trolltech Qt libraries\footnote{\href{http://trolltech.com/products/qt}{http://trolltech.com/products/qt}} which provided I/O access to the tests and a graphical user interface to load tests, create plots and perform some simple statistical analysis on the data. A separate C library, \emph{Lama Link}, provided a bridge between Lama View and IDL. Lama Link implements all the CDI functions as \texttt{lama\_get\_sky\_x} by sending such commands through a socket to Lama View and converting the answer into an IDL object. Lama Link functions run as dedicated threads, therefore allowing the GUI to be responsive even during long computations. Fig.~\ref{fig:dataAccess} shows how Lama View, Lama Link and the CDI work together to allow the LIFE analysis modules to access RAA test data.

In order to improve data access speed and minimize memory usage, Lama View implements \emph{lazy loading} of the RAA data. When a test is loaded, LAMA builds an index of all the files saved under that test, but it actually loads no data. It is only when an IDL command requesting specific data is called (e.g.\ \texttt{lama\_get\_sky\_y}) that LAMA accesses the test files. If the required AUX data do not exist, then LAMA downsamples the scientific data on the fly and saves these data as an additional FITS file in the test directory. This way, subsequent LAMA sessions will load that AUX file instead of downsampling the data again.

The combination of using C/C++ code to perform FITS file I/O and downsampling and the usage of lazy loading allowed LAMA to achieve an acceptable responsivity during the RAA tests: a few seconds were typically required to load and display the interesting data for the test under analysis\footnote{Consider that during such tasks LAMA is reading a few GB of data split into thousands of FITS files. For comparison, a first release of LAMA coded entirely in IDL required several minutes only to scan the FITS file headers.}.


\begin{table}[tbf]
  \centering
  \begin{tabular}{lccp{8cm}}
    \hline
    Name & Batch & GUI & Purpose \\
    \hline

    Bscope & X & & Quick-look tool for radiometric biases and currents \\

    FFT  & X & X & Spectrum analysis and noise estimation of detector output \cite{2009_LFI_cal_R2} \\

    LinG & X & X & Determination of the radiometric linearity coefficient, noise temperature and gain \\

    OCA & X &   & Simulation of the on-board data processing and analysis of
    scientific packets and row data \cite{2009_LFI_cal_D2} \\

    Offset & X &  & Study of the application of an offset before the digitization of the radiometric signal \\

    Reverie & X & & Estimation of the quantization error performed by the REBA
    packet compressor \cite{2009_LFI_cal_D2} \\

    SPR  &   & X & Estimation of the receiver bandpass \\

    Susc & X & X & Susceptibility of the detector output to external impulses \\

    THF  & X &   & Estimation of thermal transfer functions \cite{2009_LFI_cal_T3} \\

    Tun  & X &   & Tuning of the active components of the radiometers (amplifiers, phase switches) \\

    \hline
  \end{tabular}
  \label{tbl:LIFEmodules}
  \caption{A list of the analysis modules that have been implemented in
    LIFE~3.0 and used during the RCA/RAA tests. For each module the table
    reports the presence of a Graphical User Interface (GUI) and of
    a command line interface via the IDL prompt. A few of them (e.g.
    FFT) can be used in both modes.}
\end{table}


\section{Current Work on LIFE}
\label{sec:Pegaso}

During flight operation the FITS files format will be different from that used during ground tests. Moreover, people inside the DPC will not need to read data through FITS files, as they will directly connect to the flight database. It is therefore necessary to have a new I/O layer alongside RaNA and LAMA. It is also necessary to have a tool that produces the Daily Quality Reports (DQR) and the Weekly Health Reports (WHR), used to check the sanity of the instrument. For those reasons we implemented Pegaso, the LIFE module to be used during flight operations.

Pegaso uses the Lama codebase for the I/O and for the GUI with some obvious differences, e.g. flight data are not categorized in different ``tests'' but rather are a continuous flow of information coming from the spacecraft. Pegaso allows the user to access radiometric data by specifying the time range over which to load data. Pegaso uses a socket layer similar to the one used by LAMA to communicate with the analysis modules.

The new I/O interface for database data uses the MPA-DMC\footnote{\texttt{ftp://ftp.rssd.esa.int/pub/PLANCK/documentation/MPA-DMC/v02-03-01/Common/doc/index.html}} interface with the LFI database to query data by time range. The Pegaso tree is designed to follow the data structure defined by the DDL (Data Definition Layer), which defines object types in the LFI database using an XML format.

The DQR uses the socket interface and some of the LIFE modules to produce the daily report. A new XML parser is implemented in IDL to read the DQRs and to produce the WHR.


\section{Conclusions}
\label{sec:Conclusions}

LIFE has been the tool used to analyze the tests on the \Planck{} LFI instrument, starting from the radiometric (RCA) tests. The main driving goals have been the ability to interface with IDL and to provide a basic graphical interface for some common operations (such as plot production), as well as the ability to extend it for its use during further tests (RAA, \Planck{} system integrated tests) and even during flight.

These goals have been achieved by developing two basic libraries that create links between the many modules composing the software: the Common Data Interface (CDI) and Lama Link (and Pegaso Link). The CDI allows analysis modules to access radiometric data acquired during a test independently from the instrumentation tested (RCA or RAA). The Lama/Pegaso Link library integrates two C++ programs within IDL, Lama View and Pegaso View, that provide extremely fast access to RAA and in-flight data respectively. The coupling of these two libraries allows LIFE to exhibit an extreme versatility together with no compromises in performances. Both features have been important for the LFI scientific team to perform all the required analysis on the LFI instrument.

LIFE is going to play a key role during the flight of \Planck{}, due to the fact that the new LIFE module being developed (Pegaso) will be used to produce the daily and weekly reports containing the status of the LFI instrument. The success in the modular, mixed-language (IDL/C++) approach used in the development of Lama made the developer choose it as the baseline for Pegaso too. A first release of Pegaso has been delivered in 2008 and validated during the System Operation Validation Tests (phase 2) at the LFI DPC in Trieste.

\section*{Acknowledgements}

This paper was created under ASI contract Planck LFI Activity of Phase E2.
The US Planck Project is supported by the NASA Science Mission Directorate.

\begin{appendix}

\section{Use of downsampled data to estimate the statistical properties of a signal}
\label{sec:downsampledMath}

Both RaNA and Lama use downsampled data streams (called ``AUX streams'') for quick computations and plots. In this appendix we shall show the mathematical formulae used by the two environments to create AUX streams and to estimate the statistical properties of the full signal from them.

\subsection{Creation of the AUX data stream}

The AUX data stream is a $4\times N$ matrix where each row is composed of four heterogeneous elements $(t_i, n_i, \bar x_i, \sigma_i)$. Each point is calculated from the set $S_i$ of those $n_i$ points of the full data stream whose time falls into the interval $[t_i, t_i + 1\, \mathrm{s}]$ and whose mean and standard deviation are computed using the standard definitions:
\begin{equation}
\label{eq:AUX}
\bar x_i = \frac{1}{n_i} \sum_{k \in S_i} x_k, \quad \sigma_i = \sqrt{\frac{\sum_{k \in S_i} \bigl(x_k - \bar x_i \bigr)^2}{n_i - 1}},
\end{equation}
where $x_k$ is the $k$-th element of the full data radiometric stream. This matrix is computed once and saved into a FITS file, for speed purposes.

\subsection{Statistics from the AUX samples}

Once the AUX data stream has been created, it can be used to determine the average and the standard deviation of any subset of points $S$ of the \emph{full} data stream, provided that $S$ can be written\footnote{To satisfy this requirement, LIFE always uses sets containing points in the time window $[t_0, t_1]$, where both $t_0$ and $t_1$ are rounded to the nearest integer. This means that the time resolution of these statistics is always 1\ second.} as the union of $N$ sets $S_i$. The formulae for the mean and the standard deviation of the full data stream are computed using the following formulae:
\begin{equation}
\bar x_S = \frac{1}{N} \sum_i \bar x_i, \quad
\sigma_S = \sqrt{\frac{\sum_i \bigl( (n_i - 1) \sigma_i^2 + n_i \bar x_i^2 \bigr) - N \left( \sum_i n_i \bar x_i \right)^2}{N - 1}}.
\end{equation}
By substituting $x_i$ and $\sigma_i$ with the definitions in eq.~\ref{eq:AUX} it can be shown that these equations lead to the correct estimation of the mean value and the standard deviation of the samples in the set $S$.

\end{appendix}


\bibliographystyle{plainnat}
\bibliography{off-line-analysis,references_prelaunch_forJI}

\end{document}